\documentclass[conference]{IEEEtran}

\IEEEoverridecommandlockouts

\usepackage{cite}
\usepackage{amsmath,amssymb,amsfonts}
\usepackage{algorithmic}
\usepackage{multirow}
\usepackage{graphicx}
\usepackage{textcomp}
\usepackage{xcolor}
\usepackage{comment}
\usepackage{subfig}
\usepackage{wrapfig}
\usepackage{placeins}
\usepackage{float}
\usepackage{amsthm}

\usepackage{scalerel,stackengine,amsmath}
\usepackage[ruled,linesnumbered]{algorithm2e}
\SetKwFor{For}{for (}{) $\lbrace$}{$\rbrace$}

\newcommand\equalhat{\mathrel{\stackon[1.5pt]{=}{\stretchto{%
    \scalerel*[\widthof{=}]{\wedge}{\rule{1ex}{3ex}}}{0.5ex}}}}

\def\BibTeX{{\rm B\kern-.05em{\sc i\kern-.025em b}\kern-.08em
    T\kern-.1667em\lower.7ex\hbox{E}\kern-.125emX}}

\newtheorem{definition}{Definition}[section]

\usepackage{algorithm2e}

\SetCommentSty{mycommfont}

\IEEEoverridecommandlockouts

\begin{document}

\title{A Two-phase Metamorphic Approach for Testing Industrial Control Systems
\thanks{This work has received funding from Horizon 2020 programme under grant agreement No. 957212 - VeriDevOps project.}}

 \author{\IEEEauthorblockN{Gaadha Sudheerbabu\IEEEauthorrefmark{1},
 Tanwir Ahmad\IEEEauthorrefmark{1}, Filip Sebek\IEEEauthorrefmark{3}, Dragos Truscan\IEEEauthorrefmark{1}, Jüri Vain\IEEEauthorrefmark{1}\IEEEauthorrefmark{2} and
 Ivan Porres\IEEEauthorrefmark{1}
 }
 \IEEEauthorblockA{
 \IEEEauthorrefmark{1} Dept. of Information Technology,
  Åbo Akademi University, Turku, Finland, Email: firstname.lastname@abo.fi, \\
  \IEEEauthorrefmark{2} High-assurance Software Laboratory, Tallinn Technical University, Tallinn, Estonia, Email: juri.vain@ttu.ee\\
\IEEEauthorrefmark{3} ABB, Sweden, Email: filip.sebek@se.abb.com}
 }

\maketitle

\begin{abstract}

We elaborate on a metamorphic approach for testing industrial control systems. The proposed approach consists of two phases: an exploration phase in which we learn about fault patterns of the system under test and an exploitation phase where the observed fault patterns are used for targeted testing. Our method extracts metamorphic relations and input space of the system from its requirements. The seed input used for testing is extracted from the execution logs of the system and used to generate source tests and follow-up tests automatically. The morphed input is constructed based on the seed input and refined using a set of constraints. The approach is exemplified on a position control system and the results show that it is effective in discovering faults with an increased level of automation.
\end{abstract}


\section{Introduction}

The functional correctness of industrial software systems is of utmost importance as a system failure may incur significant financial or even human life losses. Testing of such industrial systems are further complicated due to the lack of a test oracle~\cite{weyuker1982testing}. Metamorphic testing (MT) was introduced by Chen et al.~\cite{chen2020metamorphic} as a solution to test systems when the expected output or test oracle of the \textit{system under test} (SUT) is not available to compare the actual output of the SUT against its expected output. In MT, the behavioural or functional properties of the system are defined using generic relations known as \emph{metamorphic relations (MRs)} between different sets of inputs and their expected outputs.  These relations are used to verify functional correctness instead of mapping specific inputs to their expected outputs.

However, the recent surveys on MT highlight several questions remain open for further investigation: how to create the MRs, how to define the follow-up test cases, and how to automate different phases of the process. In this paper, we propose a two-phase MT approach to circumvent the need for a traditional pre-defined test oracle:  
a) \textbf{exploration}: extracting the MRs from system specifications, generating source test cases automatically from real-execution data by analyzing data logs, and generating follow-up test cases automatically from source test cases, and b) \textbf{exploitation}: identifying fault patterns via random test generation and exploring them in more detail via guided test generation.    
In our approach the definition of MRs is manual based on domain expert knowledge, but the test generation, execution, and verdict assignment are automatic. In fact, studies have shown that manual testing with domain expert involvement can be more effective than fully automated testing \cite{DBLP:conf/ast/ZafarAE22}. We exemplify our approach on a position control system that determines the position of a hanging load attached to the hoisting frame of a crane.

\section{Overview of the approach}

Metamorphic testing performs as follows \cite{liu2012new}: a) identify MRs based on system properties defined in the software specification, b) generate a \emph{source test case} passing the seed input to the system, c) generate \emph{follow-up test cases} from the \emph{source test case} based on MRs and execute them, and d) compare the results of \emph{source and follow-up test cases} if they  satisfy the MR.

An MR is composed of two parts: an \emph{input relation} and \emph{output relation} \cite{liu2012new}. An input relation represents the relation between the inputs of source and follow-up test cases, whereas an output relation represents the relation between the outputs of the source and follow-up test cases. A \textit{source test case} is the first set of tests performed using \emph{seed inputs}. The seed inputs are transformed into \emph{morphed inputs}. The \textit{follow-up test cases} are performed using these \emph{morphed inputs}. In addition, an \textit{implication} between the outputs of {source and follow-up test cases} is needed to specify the impact of input transformations on their corresponding outputs. 
Chen et al. \cite{chen2018metamorphic} presented the MT methodology and defined MR as follows:

\begin{definition}
\textbf{(Metamorphic relation):} Let \textit{f} be a target function or algorithm. An MR is a necessary property of \textit{f} over a sequence of two or more inputs $ \langle x_1, x_2, ..., x_n \rangle $ where $n \geq 2$, and their corresponding outputs $ \langle f(x_1), f(x_2),..., f(x_n) \rangle $. It can be expressed as a relation $ \mathbf{R} \subseteq X^n \times Y^n $, where $X^n$ and $Y^n$ are the Cartesian products of \textit{n} input and \textit{n} output spaces, respectively. 
\label{def:MR_Chen}
\end{definition}

We extend the above definition, by refining $\mathbf{R}$ into $R_{in}$ and $R_{out}$, where the satisfiability of MR output relation $R_{out}$ by outputs $Y_s$ and $Y_m$ presumes also that their corresponding morphed inputs $X_m$ satisfy respectively MR input relation $R_{in}$. That is, 
given $f(x_i) = y_i$ and $f(x_j) = y_j \forall (x_i,x_j)$, then $R_{in}(x_i, x_j) \implies R_{out}(y_i, y_j)$, where
$f$ denotes the function that creates outputs $(y_i, y_j)$ in response to inputs $(x_i, x_j)$, $R_{in}$ is input MR and $R_{out}$ is output MR.

Concretely, for a given sequence of inputs $X^c \subseteq X$ under a constraint $C$, an MR \textit{R} should hold for any corresponding output of the system, that is $f(X^c_s) \mathbf{R} f(X^c_m)$, where $X^c_s, X^c_m \subseteq X^c$. Furthermore, we consider $\mathbf{R}$ to be of any of the types defined in \cite{segura2017metamorphic}: equivalence, equality, subset, disjoint, complete, and difference. 

\begin{figure}[t]
\centering
  \includegraphics[width=1\linewidth]{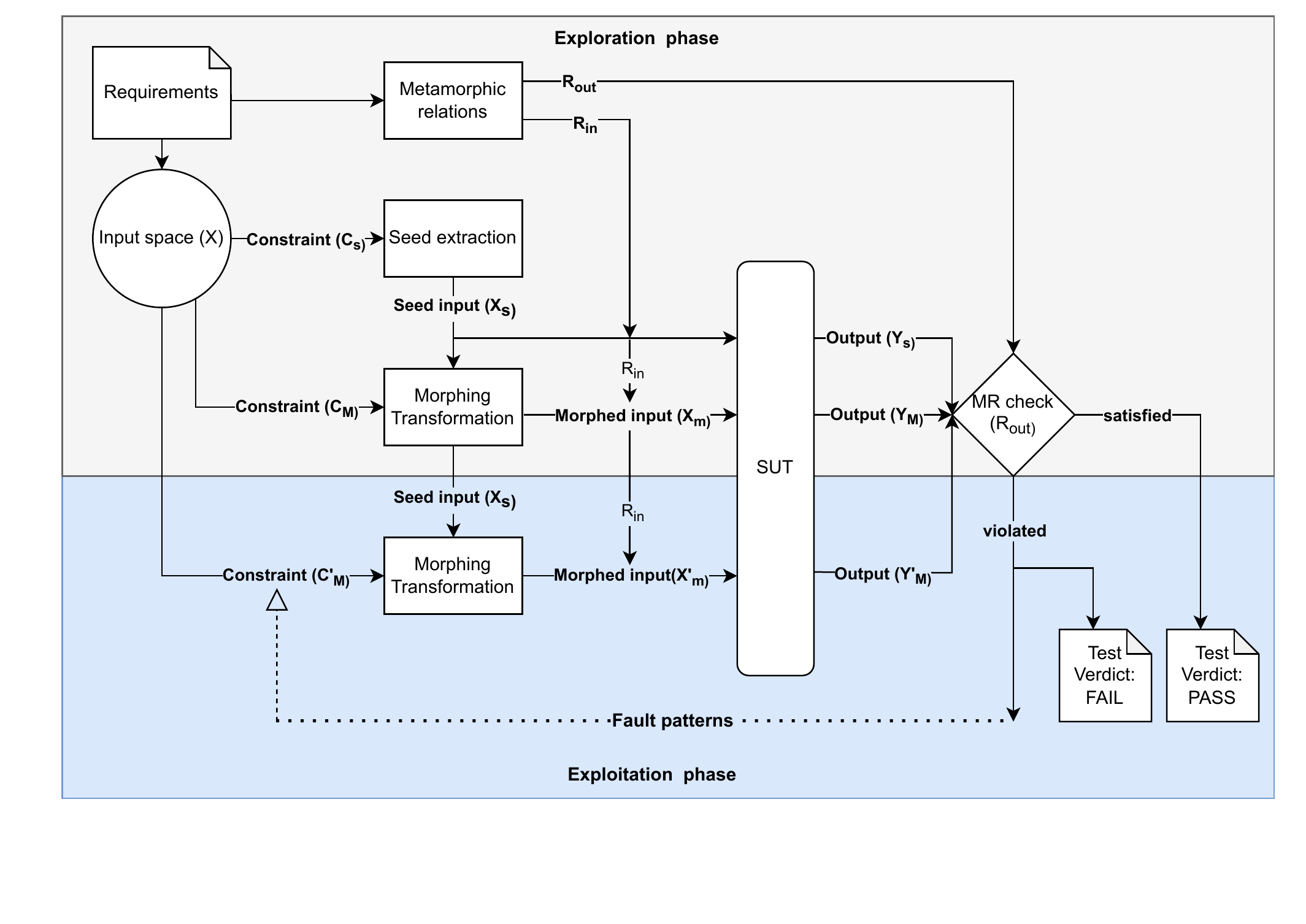}
  \caption{Overview of the metamorphic testing approach }   \label{fig:fig2mtapproach}
\end{figure}

Our MT approach is shown in Figure~\ref{fig:fig2mtapproach}. 
It is applied using two phases: \textit{exploration phase} and \textit{exploitation phase}. In the exploration phase, $X_s, X_m$ are extracted/created from X satisfying a set of constraints $C_s, C_m$ specific to the SUT. Then $X_s, X_m$ are executed against the SUT and the corresponding \textbf{seed output} $Y_s$ and respectively \textbf{morphed output} $Y_m$ are collected and satisfiability of $R_{out}(Y_s,Y_m)$ is checked, where $ Y_s =f(X_s)$ and $Y_m =f(X_m)$.

From those pairs of seed and morphed inputs $(X_{s_i}, X_{m_i})$ which fail the initial MR, we manually extract, in the exploitation phase, \textbf{fault-inducing patterns} of the input space. Based on them, we define  $C_m'$ as a more restrictive constraint to be satisfied by morphed inputs
$X_m'$ which we use to verify the output metamorphic relation $R_{out}(Y_s,Y_m')$, where $ Y_s =f(X_s)$ and ${Y_m'} = f({X_m'})$.

The novelty of our approach stands in the fact that $C_m'$ allows us to define a refined morphed input that tests the system with more precision and effectiveness, by focusing the testing on the parts of the input with a higher probability of discovering faults.

\subsection{Running Example}

\begin{wrapfigure}{r}{0.4\linewidth}
 \vspace{-1pc}
\centering
  \includegraphics[width=0.6\linewidth]{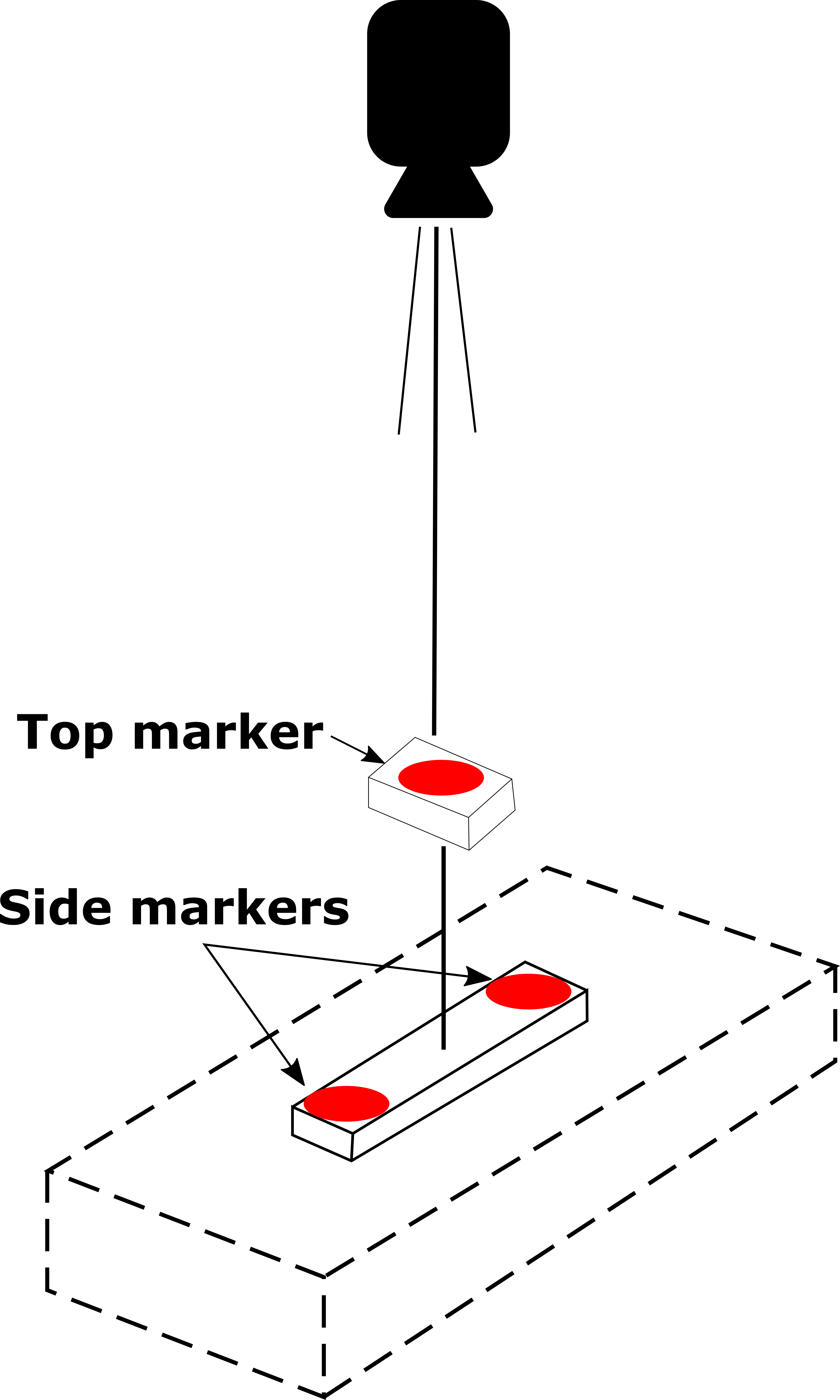}
  \caption{Positional markers in PCS}   \label{fig:pcs}
  \vspace{-1pc}
\end{wrapfigure}

The ICS is a \textit{Position Control System} (PCS) which determines the position of a hanging load using attached markers on the hoisting frame. The PCS regularly receives up to 26 markers as [x,y] pixel coordinates from a camera module. The input may contain three markers on the hoisting frame attached to the  load, as well as different light reflections in the environment (water, rain, snow, dust, etc.) which the camera filter was not able to remove. Only the markers corresponding to the three markers placed on the hoisting frame carrying the load are the \emph{true markers} that determine the position of the load (see Figure~\ref{fig:pcs}). The two markers placed on the sides of the hoisting frame are referred to as \emph{side markers}. The \emph{top marker} is used to detect the tilt of load and to increase the probability for the find algorithm to identify the true markers. 

For each set of markers, the PCS tries to identify the true markers and to discard the markers corresponding to reflections. The PCS produces two outputs: a Boolean value \textit{found} indicating whether  \textit{true markers} is identified and a vector of three integers \textit{[$I_{tm_1},I_{tm_2}, I_{tm_3}$]}, indicating the index in the input marker array of the positional markers identified as \textit{true markers}. Whenever the PCS is not able to identify the true markers consistently, the entire system can potentially move to an unsafe state and requires human intervention.  
In the above context, we define the output of the PCS as follows: 

\begin{definition}
The \textbf{output} of the Position Control System, $f(\{m_1, m_2,...,m_n\})$ is a pair $(found, [I_{tm_1},I_{tm_2}, I_{tm_3}])$, where $m_i$ and $tm_i$ are positional markers with two coordinates $x_i$ and $y_i$, $\{tm_1,tm_2, tm_3\} \subseteq \{m_1, m_2,...,m_n\}$,  $found = TRUE | FALSE$ and $[I_{tm_1},I_{tm_2}, I_{tm_3}]$ is the vector of indexes of true markers provided that $found = TRUE$. 
\label{def_outPCS}
\end{definition}

\subsection{Metamorphic relation}

In our approach, we extract the MR from the requirements of the SUT as follows: 
"\textit{Assuming that the system is able to classify correctly a set of markers received from camera module in the absence of reflections (noise), the system should be able to classify correctly the same inputs in the presence of reflections}". 

This can be formulated as the following metamorphic relation: $f(X_s) \equiv f(X_s \cup X_n)$.

\subsection{Creating the seed input} 
In our approach, we choose the seed input as a series of true marker triplets $ X_s = \{s_1, s_2, \dots, s_k\}$, where $s_i = \{tm_1^i,tm_2^i, tm_3^i\}$. We extract the seed input from previous executions of the PCS by extracting those log entries that only contain three positional markers and which were classified correctly as true markers.  The spatial continuity of the image coordinates is validated via checking the distance between consecutive image coordinate values against a predefined allowed range of movement. When the seed input data set is extracted from the execution trace, we run an initial test session against the SUT to confirm that all the input marker positions are classified correctly. In case execution logs are not available, the seed input data can be collected from the simulation environment of the PCS that is validated against a real crane for the set of inputs the seed is extracted from.

\subsection{Creating the morphed input}

In our case, the morphing transformation takes each sample in the seed input $X_s$ and adds markers corresponding to reflections, which we denote as \textit{noise}.

\begin{definition}
\label{def:noise}
\textbf{(Noise)}: A series of noise markers corresponding to environment reflections $X_n \equalhat \{n_1,n_2,n_3,...,n_{j} \} $, where $n_i \in X$.
\end{definition}

In the \textbf{exploration phase}, we use random generated noise to perform an initial exploration of the SUT in order to collect observations and identify fault patterns. 

To this extent, we create random noise coordinate pairs of marker vectors of different lengths ranging from 1 to 23. These noise vectors are appended to the seed input one at a time. The algorithm for generating morphed input generates random (x, y) coordinates with a value in range [0,131072], which is the size of the camera frame.

\begin{definition}
\label{def:morphedinput}
\textbf{(Morphed input)}: A series of markers  
$X_m$ $ = \{m_1, m_2,..., m_k \}$, 
where each sample $m_i = \{tm_1^i, tm_2^i, tm_3^i ,n_1^i,n_2^i,n_3^i,...,n_{j}^i \}$ with $ j \leq 23 $, is the combination of seed input markers $X_s$ and noise markers $X_n$.
\end{definition}

In the \textbf{exploitation phase}, we analyze noise patterns in the \textit{morphed input} that caused the system to make incorrect classifications. This led to the following observations: \textit{the set of markers containing two} or \textit{three noise markers having the same geometrical pattern of true markers can trigger faulty behavior of the system}. Therefore, we refine the morphed input to a more constrained version of the input space to exploit the above mentioned fault patterns. 

\begin{definition}
\label{def:refinedmorphedinput}
\textbf{(Refined Morphed input)}: A series of markers samples
${X_m'}= \{m_1, m_2,..., m_k \}$, where each sample 
$ m_i = \{tm_1^i, tm_2^i, tm_3^i ,n_1^i,n_2^i \} $ in the first follow-up test and 
$m_i = \{tm_1^i, tm_2^i, tm_3^i ,n_1^i,n_2^i,n_3^i\} $ in the second follow-up test. $C_m'$ is the restrictive constraint used to refine the added noise to two and three noise markers. The series ${X_m'}$ is the combination of seed input markers $X_s$ and restricted set of noise markers $X_n'$ generated using the constraint $C_m'$. 
\end{definition}

In order to automate the creation of the noise markers, we create replicas of the true markers, thus obtaining  a similar geometrical pattern in the noise. For each sample of true  markers in the seed input we distribute the noise markers in a rectangular grid pattern in the camera frame, in order to obtain a uniform sampling of the input space. 

\begin{wrapfigure}{r}{0.5\linewidth}
\vspace{-1pc}
  \centering
     \includegraphics[width=1.1\linewidth]{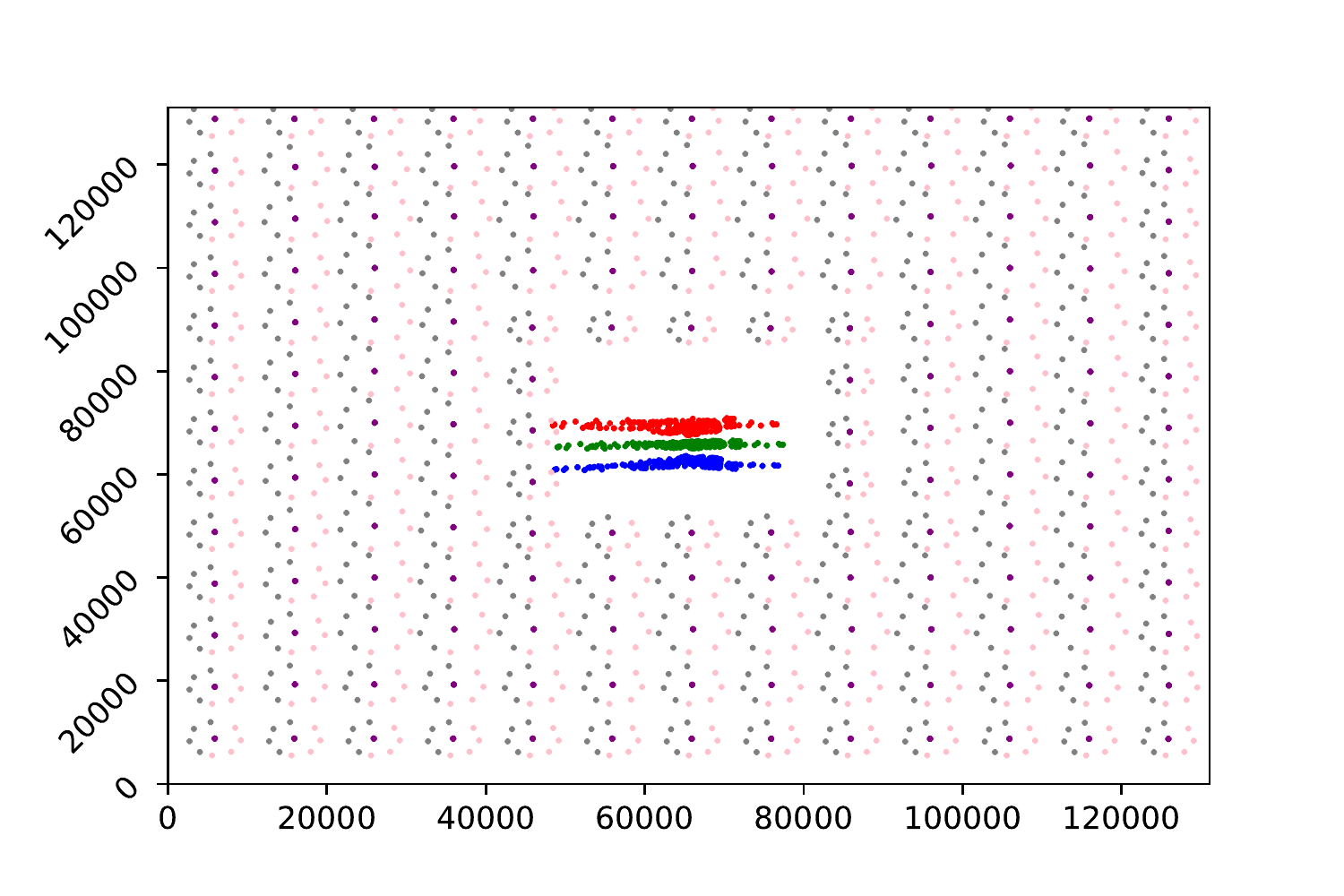}      
     \caption{Test data distribution for the guided star approach}
     \label{fig:GuidedDataDistribution}
     \vspace{-1pc}
\end{wrapfigure}
In addition, each replica of true marker pair placed on the grid is rotated by angles $45^0$, $90^0$, and $135^0$ to distribute the noise markers in a star like pattern (see Figure~\ref{fig:GuidedDataDistribution}). We note that, the approach is completely automated and it allows us to change the number of generated tests by changing the density of the grid and number of rotations of the noise markers.

\subsection{Test execution}

The tests are generated in offline mode. Test execution is performed using the \texttt{Pytest} testing framework \cite{hunt2019pytest} by setting up an adapter to connect to the SUT. Open Platform Communications Unified Architecture (OPC UA) \cite{leitner2006opc} is used as the adapter to connect to the CODESYS development environment where the PLC application program resides. The test suite contains the functions to set up OPC server-client connection. It also has a one-time setup to read the input from an input file, send it to SUT, and collect the execution results to verify if the MR defined between the source and follow-up test output holds.

\section{Experiments and evaluation}
\label{sec:experiments}

For all experiments used in this section, we use a seed input with 625 samples, each containing a sequence of three true markers extracted from execution logs. Each entry in the seed input is verified first that it is classified correctly as the true markers by the system. 

Based on the outcomes of the executed follow up test cases we can categorize the test cases as follows: \textit{true positive} (TP) -- the system identifies the actual positional markers as true markers - expected behavior, \textit{false positive} (FP) -- the system identifies reflection markers as true markers - unexpected behavior, \textit{false negative} (FN) -- the system fails to identify the true markers even if they were present in the input - unexpected behavior, and \textit{true negative} (TN) -- the system does not identify true markers when the input does not contain true markers. This is not applicable since in our approach since the test input always contain true markers.

The TP classification of true markers in the morphed output satisfies the MR and counts as tests that do not fail. The failed tests include FPs, which indicate an incorrect identification and FNs, which indicate missed identification of true markers placed in the first three positions in the morphed input.

\begin{figure*}[!t]
     \centering
          \subfloat[Exploitation: FPs for markers=5]{
           \includegraphics[width=.33\linewidth]{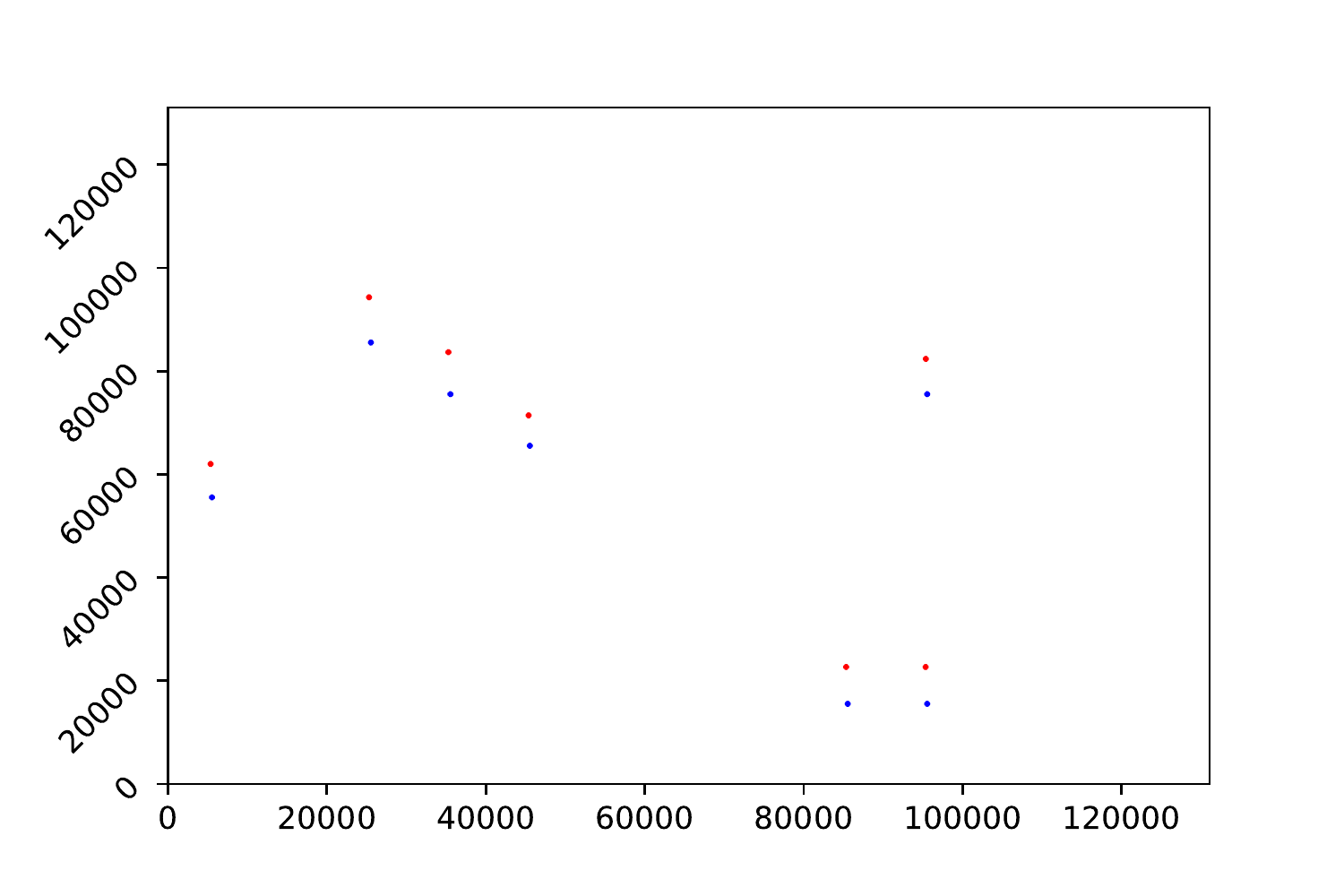}}
           \subfloat[Exploitation: FNs for markers=5]{
        \includegraphics[width=.33\linewidth]{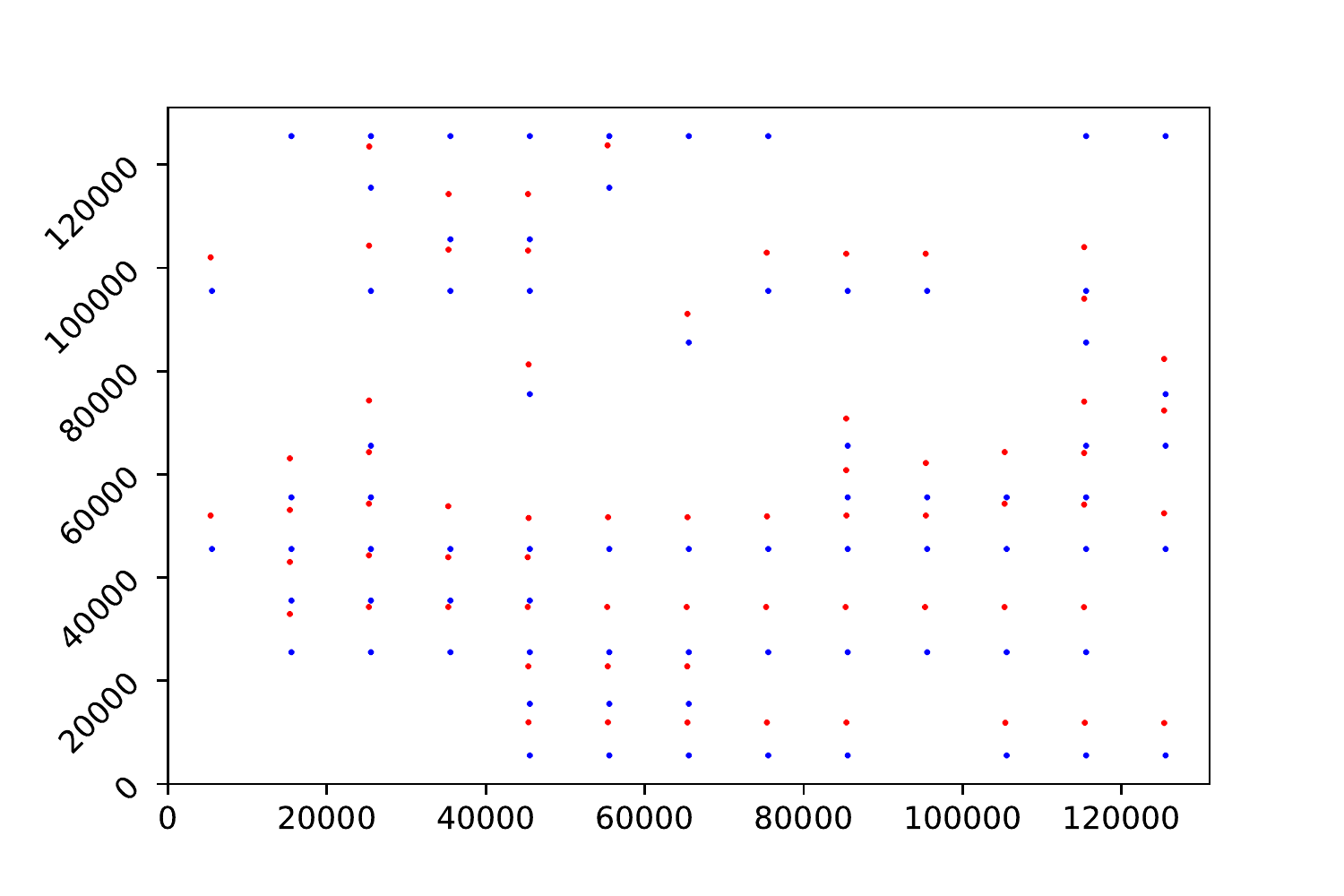}          } 
     \subfloat[Exploitation: FPs for markers=6]{
        \includegraphics[width=.33\linewidth]{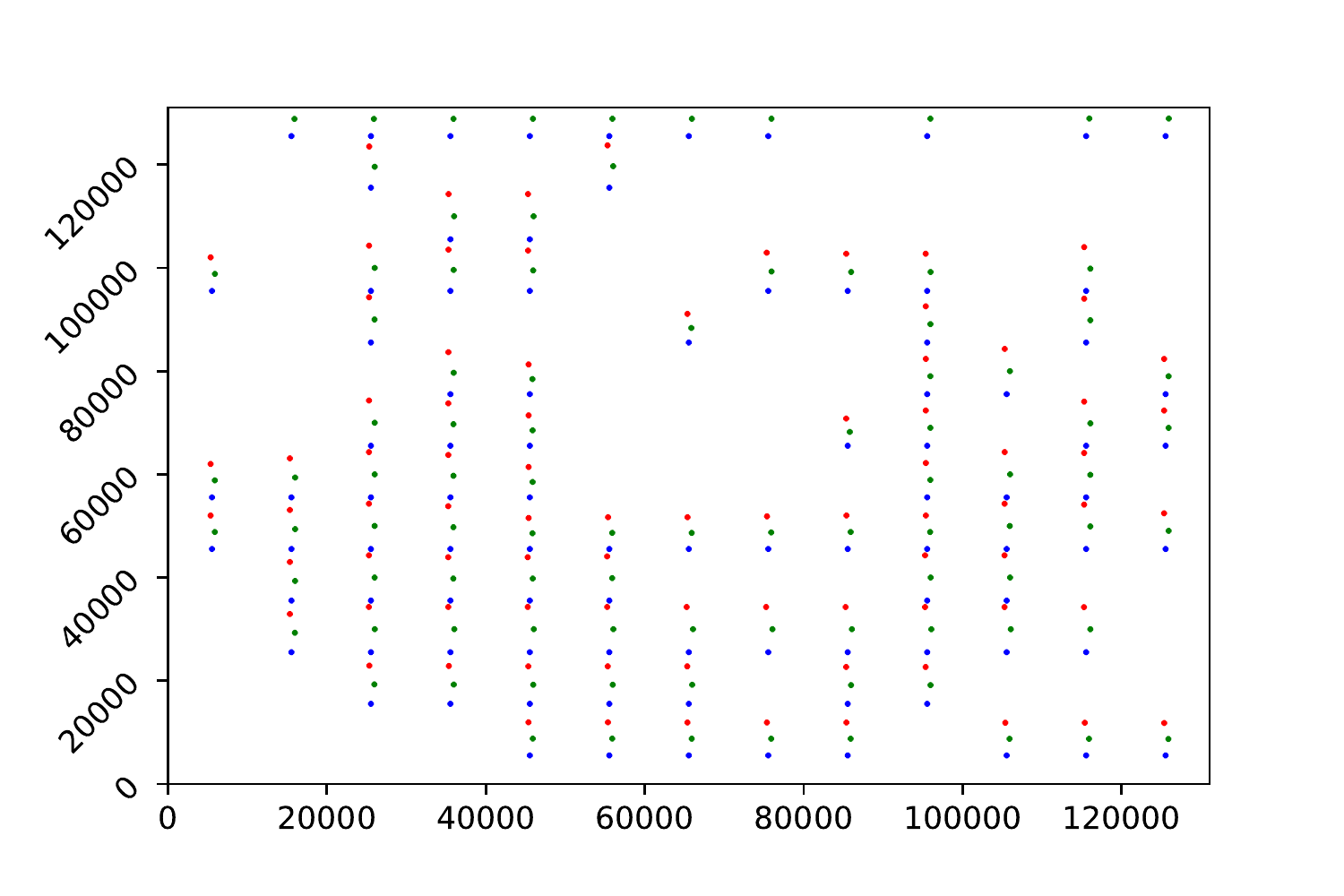}       }
     \caption{Input distribution for failed tests in the exploitation phase}
     \label{fig:FPresults}
\end{figure*}

In the exploration phase, the test generation algorithm produces $ 625 \times 10 \times 23 = 143 750 $ follow up tests. From the total of 143 750 executed tests, 143 615 satisfy the metamorphic relationship, while 135 do not. From the failed tests, 39 selected the wrong combination of inputs as true markers, whereas 96 could not find true markers among the inputs although they were present.

\begin{figure}[h]
\vspace{-1.2pc}
     \centering
     \subfloat[morphed input w.r.t FP output]{
        \includegraphics[width=0.5\linewidth]{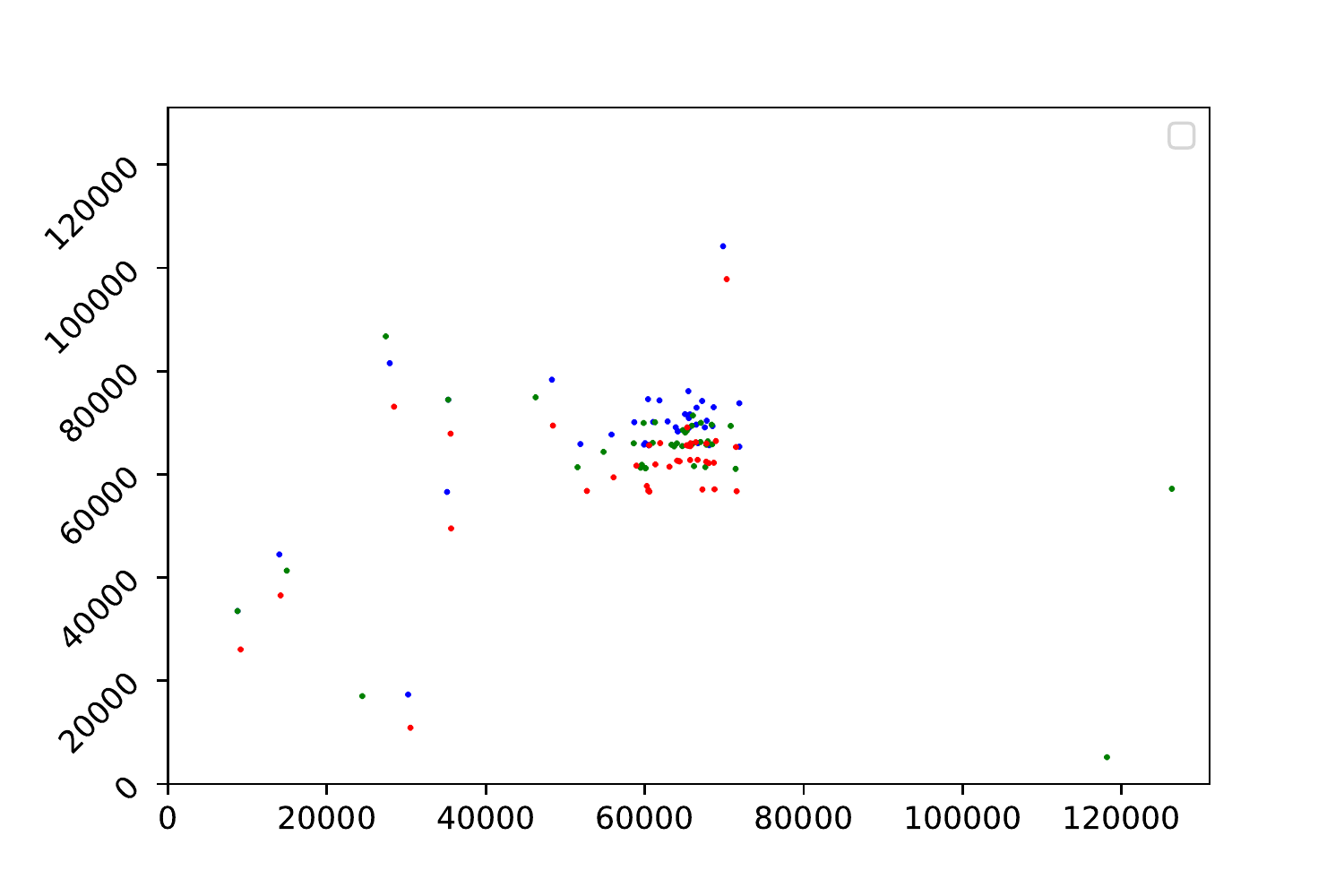}
      }
     \subfloat[morphed input w.r.t FN output]{
        \includegraphics[width=0.5\linewidth]{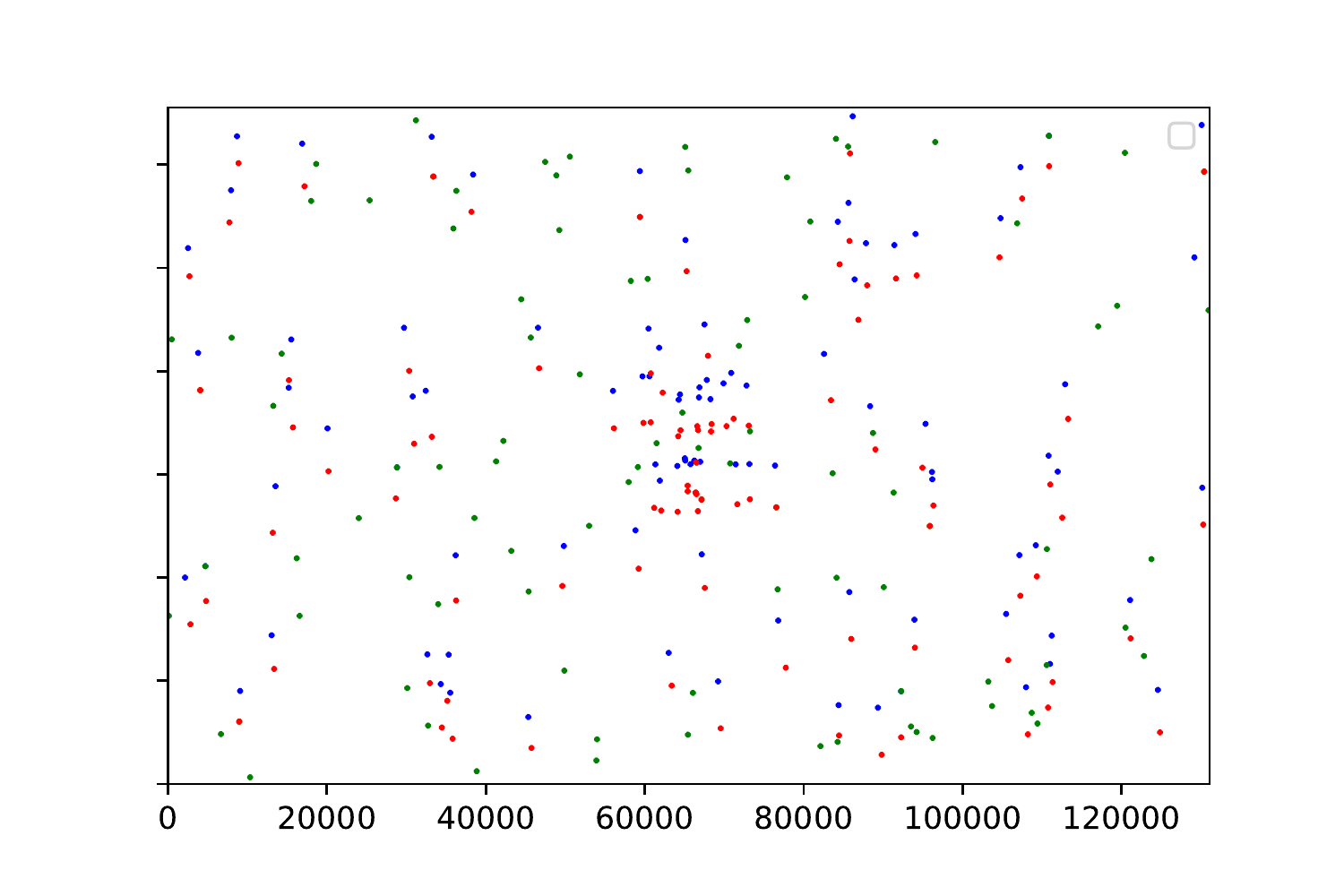}
      }
     \caption{Morphed input corresponding to incorrect classifications in the exploration phase}
     \label{fig:Rclassification}
\end{figure}

The geometric distribution of incorrectly classified data points is shown in Figure~\ref{fig:Rclassification}. Further analysis of the failed tests, provides us with the following observations. FP results occurred when the input set contained either a set of \textit{two noise markers} resembling the pattern of true side markers or a set of \textit{three noise markers} resembling the geometrical pattern of the true marker triplet. FN results occurred when the input set contained either a number of markers greater than or equal to 6 or a set of \textit{two noise markers} resembling the pattern of the true side markers. 

In the exploitation phase, we ran two separate testing sessions in which the refined morphed input contains two and three noise markers respectively, besides the true markers. In both cases, the test generation algorithm produced 2400 refined morphed test inputs. These refined morphed inputs are created by replicating and rotating the seed input markers. It results in $ 625 \times 4 = 2500 $ noise markers and discarding the samples that do not fit in the 131072 x 131072 frame  after the \textit{rotate} action. The results of the test execution are shown in Table~\ref{tab:test-exec}. For the test session with five input markers, 7 FP and 74 FN classifications are identified, whereas, for the subsequent test with six input markers, no FN and 92 FP classifications are identified. The distribution of the noise markers in the input space for failed tests in  the exploitation phase is shown in Figure~\ref{fig:FPresults}. 
The test execution results of the guided method where the number of markers is 5 contain more FNs indicating that the system is not identifying the true markers when a replica of two side markers is added as noise. However, the FP results of follow-up test where the number of markers is 6 reveal that a replica of 3 true markers can trigger an incorrect identification and compromise the functional safety of the system. It is also observed that a replica of two side markers has low chances of causing FPs when compared to the noise created with a replica of 3 true markers. Moreover, only the noise markers corresponding to the exact replica of true markers triggered the incorrect identification of true markers. In addition, we can observe that the noise markers rotated by angles $45^0,90^0,135^0$ resulted in TP test cases, where the system correctly identified the true markers despite the presence of noise. 

Table \ref{tab:test-exec} also shows the corresponding Fault Detection Ratio (FDR) \cite{segura2016survey} for each phase as the number of tests that found a fault in the entire tests suite. As expected, the exploration phase has a very low FDR due to the random test generation, whereas in the exploitation phase, FDR has increased around 33 to 44 fold.

\begin{table}[!h]
\centering
\begin{tabular}{|l|r|r|r|r|r|r|}
\hline
Method                                & No. of  & No. of & TPs & FPs & FNs & FDR  \\ 
                               &  markers &  tests &  &  &  &  \\ \hline
Exploration                          & 4 - 26            & 143750          & 143615         & 39              & 96              & 0.0009                \\ \hline

\multirow{2}{*}{Exploitation}       & 5                 & 2400   &  2319  &  7  & 74  &      0.03             \\ \cline{2-7} 
        & 6    & 2400   & 2308   & 92   & -  & 0.04
        \\ \hline
\end{tabular}
\caption{Test execution results}
\label{tab:test-exec}
\end{table}

\section{Discussion and Conclusions}

In this work, metamorphic testing is effectively applied to detect faulty behavior in industrial control systems. The identification of a metamorphic relation is done manually based on the specification of the system. A known challenge in the identification of MRs is the need for domain expertise to assess the expected input and output behaviour of the system. As future work, we plan to automate the identification of MR for an ICS from its specification and to explore the applicability of MT for fault localization and program repair in industrial systems. It is also of interest to combine metamorphic and mutation-based approaches for testing ICS and to apply heuristic techniques for minimization of test suites.

\end{document}